\documentclass[11pt,english]{article}
\usepackage{times}

\usepackage{babel}
\usepackage{array,color}
\usepackage{times}
\usepackage{helvet}
\usepackage{courier}
\usepackage{amsmath,amsfonts,relsize}
\usepackage{algorithm}
\usepackage{algpseudocode}
\usepackage{graphicx}

\newcommand{\tworows}[2]{$\genfrac{}{}{0pt}{}{\mbox{#1}}{\mbox{#2}}$}

\title{The Shapley Axiomatization for Values\\ in Partition Function Games}

\author{Oskar Skibski \\
University of Warsaw, Poland \!\!\!\! \\
\and 
Tomasz P. Michalak \\
University of Oxford, UK \!\!\!\! \\
\and Michael Wooldridge \\
University of Oxford, UK
}

\newtheorem{theorem}{Theorem}
\newtheorem{corollary}{Corollary}

\newcommand{\ForAllLine}[2]{%
    \State\algorithmicforall\ {#1}\ \algorithmicdo\ {#2} \algorithmicend\ \algorithmicfor%
}

\def\defeq{\stackrel{\mathrm{\mathsmaller{def}}}{=}}
\def\P{\mathcal {P}}

\begin{document}

\maketitle

\begin{abstract}
One of the long-debated issues in coalitional game theory is how to extend the Shapley value to games with externalities (partition-function games). When externalities are present, not only can a player's marginal contribution---a central notion to the Shapley value---be defined in a variety of ways, but it is also not obvious which axiomatization should be used. Consequently, a number of authors extended the Shapley value using complex and often unintuitive axiomatizations. 
Furthermore, no algorithm to approximate any extension of the Shapley value to partition-function games has been proposed to date.
Given this background, we prove in this paper that, for any well-defined measure of marginal contribution, Shapley's original four axioms imply a unique value for games with externalities. As an consequence 
of this general theorem, we show that values proposed by Macho-Stadler et al., McQuillin and Bolger can be derived from Shapley's axioms. 
Building upon our analysis of marginal contribution, we develop a general algorithm to approximate extensions of the Shapley value to games with externalities using a Monte Carlo simulation technique.
\end{abstract}


\section{Introduction}
\noindent Coalitional game theory provides a rich and compelling framework for
modeling and understanding cooperation in multi-agent
systems~\cite{Chalkiadakis:et:al:2011}. Most research on coalitional games in the computer
science literature has been devoted to \textit{characteristic function
  games}. The key component of such games is a characteristic
function, which assigns to every subset (or \emph{coalition}) of
agents a real number indicating the value that the coalition could
obtain through cooperation. An explicit assumption in this model is
that the value of any coalition is independent of how other agents
choose to cooperate. However, in many realistic settings, this
assumption clearly does not hold.  For example, an alliance of
smartphone producers who successfully convince clients to adopt their
operating system can increase their market position to the detriment
of their competitors: this is an example of a \emph{negative
  externality}. Positive externalities are also possible, for example
when one coalition achieves a goal that subsumes the goal of another
coalition. A key issue is then how to represent games that exhibit
such externalities, and for this purpose, \emph{partition function
  games} were introduced by Lucas and Thrall
\cite{ThrallLucas63}. In these games, the value of a
coalition is assumed to depend not just on the composition of that
coalition, but on the other coalitions that co-exist with it.
Partition function games have recently received considerable attention
in the computer science community
\cite{Rahwan:et:al:2012,Banerjee:Landon,Michalak10amas,Ichimura:et:al:2011,deKeijzer:Schafer:2012}, and will be
the focus of this paper.

A key question in coalitional game theory is how the payoff that a
coalition obtains through cooperation should be divided among
coalition members. Many payoff division schemes have been proposed,
of which the best known and most influential is due to
Shapley~\cite{Shapley53}. Shapley argued that, in order to ensure
fairness, each agent should obtain a share in the payoff that is
proportional to its \textit{marginal contribution to the game}, and
this should be computed as a function of agents' \textit{marginal
  contributions} to individual coalitions.  Shapley proved that, if
the coalition of all agents (the \textit{grand coalition}) forms, and
there are no externalities, then there exists a unique payoff
distribution that satisfies the axioms: 
(i)~\emph{efficiency}---the
entire value of the grand coalition should be distributed;
(ii)~\emph{symmetry}---if an interchange of two agents does not affect the value of any coalition, then these agents are symmetric and should receive equal payoffs;
(iii)~\emph{null-player
  axiom}---an agent with no marginal contribution to any coalition should
receive no payoff; 
and (iv) \emph{additivity}---the division scheme
should be additive over games. The strength of the Shapley value
derives from the fact that, although the above axioms are simple and
intuitive, there is no other payoff division scheme for characteristic
function games that satisfies all four of them. As a matter of fact, 
a number of alternative axiomatizations again uniquely yield the Shapley value (e.g., Myerson's \cite{Myerson80} balanced-contribution axiomatization, Young's \cite{Young85} monotonicity axiomatization, Hart \& Mas-Colell's \cite{HartMasColell89} potential-function axiomatization).
Unfortunately, for
many classes of cooperative games, computing the Shapley value is a
computationally intractable problem, and considerable efforts have
therefore been devoted to developing efficient
heuristics or approximation algorithms~\cite{Chalkiadakis:et:al:2011}. 

Given the significance of the Shapley value, it is very natural to ask
how it might be adapted or adjusted to partition function games. However, when
externalities are present, the central notion of the marginal
contribution can be defined in a variety
of ways. As a result, to date, a number of marginality measures
(hereafter, ``marginalities'') have been proposed in the literature.
This proliferation of marginalities raises an interesting question:
for any given way of measuring marginal contribution, which axioms are
required to obtain a unique value for the game? To date, the only general answer to this question has been given by Fujinaka~\cite{Fujinaka04} for Young's monotonicity axiomatization. However, for most
marginalities it has been unknown how to obtain a unique extension of the
Shapley value to games with externalities using the four original
axioms, or whether this was even possible. This
situation has led to a range of new and increasingly exotic axioms being
proposed. However, these new axiomatizations tend to be complex,
losing the simplicity and intuitiveness of Shapley's original four
axioms. Furthermore, some axiomatizations are not based on
marginality, and for these, it is unclear how to approximate a resulting value. In fact, no algorithm to approximate any extension of the Shapley value to games with
externalities has been proposed to date.

Given this background, our contributions are twofold:
\begin{itemize}
\item In Section 3 we prove that, given any well-defined\footnote{\footnotesize See Section 2 for more details.} marginal
  contribution, Shapley's original four axioms in fact imply a
  unique value for games with externalities (an \textit{extended Shapley value}). With this general theorem, we
show that not only can values proposed by Macho-Stadler \hspace{-0.015cm}\emph{et \hspace{-0.015cm}al.}\hspace{-0.005cm}\cite{MachoStadlerEtAl07},\hspace{-0.01cm} McQuillin\hspace{-0.001cm}\cite{McQuillin09} and Bolger \cite{Bolger89} be derived
from Shapley's axioms, but that Shapley's axiomatization is equivalent to Young's monotonicity axiomatization in games with externalities.
\item Based on the general formula that, given any marginality, computes the extended Shapley value, we present in Section 4 the first algorithm to
  approximate extensions of the Shapley value to games with
  externalities.
\end{itemize}

\section{Preliminary Definitions}\label{section:preliminaries}

Let $N = \{1, 2, \ldots, n\}$ be a set of players. A \emph{coalition},
$S$, is any non-empty subset of $N$. A \emph{partition}, $P$, is any
set of disjoint coalitions whose union is $N$. For technical
convenience, we will assume that $\emptyset \hspace{-0.09cm}\in \hspace{-0.09cm} P$ for every partition
$P$. A pair $(S,P)$, where $P$ is a partition of $N$ and $S\hspace{-0.09cm}\in\hspace{-0.09cm}P$, is
called an \emph{embedded coalition}. The set of all partitions and the
set of all embedded coalitions over $N$ are denoted by $\P$ and $EC$
respectively.

A \emph{game $v$} (in \emph{partition-function form}) is given by a
function that associates a real number with every embedded coalition,
i.e., $v\hspace{-0.12cm}:\hspace{-0.12cm} EC\hspace{-0.12cm} \rightarrow\hspace{-0.09cm} \mathbb{R}$. A game has \emph{externalities}
if the value of any coalition depends on the arrangement of outside
agents (i.e., there exist two partitions $P_1, P_2$ containing $S$
such that $v(S, P_1)\hspace{-0.09cm}\neq\hspace{-0.09cm}v(S, P_2)$). Otherwise, we say that the game is
\emph{without externalities}. Such games can be
represented in \emph{characteristic-function form}: $\hat{v}: 2^{N}
\rightarrow \mathbb{R}$.
As is customary in the literature, we assume that the \textit{grand
  coalition} will form (i.e., the coalition $N$ of all players). Then
the outcome (\emph{value}) of the game is a vector
that distributes the value of the grand coalition to players.


The set of all permutations of set $N$ (i.e., one-to-one mappings from $N$ to itself) will be denoted by $\Omega(N)$. Formally, $\pi \in \Omega(N)$ is a function, although as common in the combinatorics, we will sometimes identify it with an ordering: $(\pi(1), \pi(2), \ldots, \pi(|N|))$. We will denote the set of agents that appear in this order after $i$ by $C_i^\pi$, i.e., $C_i^{\pi} \defeq \{\pi(j) \mid j > \pi^{-1}(i)\}$ ($\pi^{-1}(i)$ is the place of $i$ in the order).

We use a shorthand notation for set subtraction and union operations:
$N_{-S} \defeq N \setminus S$, and $S_{+\{i\}} \defeq S \cup \{i\}$;
we often omit brackets and simply write $S_{+i}$. Finally, to model
the partition obtained by the transfer of agent $i$ to coalition $T$
in partition $P$ we introduce the following notation:
$$\tau_i^T(P) \defeq P \setminus \{P(i), T\} \cup \{P(i)_{-i}, T_{+i}\},$$
\noindent where $P(i)$ denotes $i$'s coalition in $P$.


\subsection{The Shapley value}
In coalitional games
with no externalities, the \emph{marginal contribution} of agent $i
\in S$ to coalition $S$ is defined as the difference between the value
of $S$ with and without $i$, i.e., 
$$\hat{v}(S) - \hat{v}(S \setminus \{i\}).$$
We will denote the \textit{vector of all marginal contributions} of $i$ to all coalitions $S \subseteq N \setminus \{i\}$ in game $\hat{v}$ by $mc_i(\hat{v})$. 


Now, assume that the agents leave a certain meeting point in a randomly selected order $\pi$. Agent $i$ leaves the set $C_i^{\pi}$ and decreases the value of (the coalition in) the meeting point by his marginal contribution. Then, the Shapley value of agent $i \in N$ is the expected marginal contribution of this agent over all permutations $\pi \in \Omega(N)$.\footnote{\footnotesize Note that typically the intuition behind the Shapley value is presented as a process of entering the coalition. We adopt a reverse convention as more convenient for games with externalities.} 
Formally:
$$\varphi_i(\hat{v}) = \frac{1}{|N|!} \sum_{\pi \in \Omega(N)} \hat{v}(C_i^{\pi} \cup \{i\}) - \hat{v}(C_i^{\pi}).$$


\noindent Shapley \cite{Shapley53} famously proved that this value is the only payoff distribution scheme that satisfies the following four desirable axioms (we state them below in the general form so that they can also be used for games with externalities): 
\begin{itemize}
\item \emph{Efficiency} (the entire available payoff is distributed among agents): $\sum_{i \in N} \varphi_i(v) = v(N, \{N, \emptyset\})$ for every game $v$;

%

\item \emph{Symmetry} (payoffs do not depend on the agents' names): $\varphi(\pi(v)) = \pi(\varphi)(v)$ for every game $v$ and permutation $\pi \in \Omega(N)$;\footnote{
As permutation $\pi$ is formally a mapping from $N$ to itself, a permutation of a set is an image of $S$: $\pi(S) \defeq \{\pi(i) \mid i \in S\}$ and a permutation of an embedded coalition is defined as follows: $\pi(S, P) \defeq (\pi(S), \{\pi(T) \mid T \in P\})$. Then, a permutation of a function (e.g., game $v$ or value $\varphi$) is $\pi(f)(a) \defeq f(\pi(a))$ for $f: A \rightarrow B$ and $a \in A$.
Intuitively, value of $(S, P)$ in permuted game $\pi(v)$ equals the value of an embedded coalition obtained by replacing all players $i$ from $(S, P)$ with $\pi(i)$. For example, if $\pi$ exchange $1$ and $2$, then $\pi(v)(\{\{1\}, \{2,3\}\}) = v(\{\{2\}, \{1,3\}\})$.
}
\item \emph{Additivity} (the sum of payoffs in two separate games equals the payoff in a combined game): $\varphi(v_1 + v_2) = \varphi(v_1) + \varphi(v_2)$ for all the games $v_1, v_2$ and $(v_1 + v_2)(S, P) \defeq v_1(S, P) + v_2(S, P)$;
\item \emph{Null-Player Axiom} (agents that never contribute to the
  value of any coalition, i.e., his marginal contribution vector is the zero vector, should get nothing)
: $mc_i(v) = \vec{0} \Rightarrow
  \varphi_i(v) = 0$ for every game $v$ and agent $i \in N$.
\end{itemize}


\subsection{Extension to games with externalities}\label{section:esv} Formalizing the notion of marginal
contribution is not straightforward for games with externalities. The
issue is that the change of value caused by an agent leaving a
coalition depends on the partition in which it is embedded. It is also important which coalition the agent joined.
For instance, given a three-player game $v$, consider agent $2$'s
contribution to coalition $\{1,2\}$ in partition
$\{\{1,2\},\{3\},\emptyset\}$. As the values $v(\{1\}, \{\{1\}, \{2\},
\{3\}, \emptyset\}), v(\{1\}, \{\{1\}, \{2,3\}, \emptyset\})$ may
differ, agent $2$ leaving coalition $\{1,2\}$ may result in a bigger
or smaller loss for the coalition $\{1, 2\}$.  The loss in value
caused by the transfer of $i$ from coalition $S$ embedded in partition
$P$ to the coalition $T$ can be understood as $v(S, P) - v(S_{-i},
\tau_i^T(P))$. We will refer to such a loss as an \emph{elementary
  marginal contribution}.

Now, various authors defined their marginal contributions as different
combinations of elementary marginal contributions. For instance, Pham
Do and Norde derived their externality-free value by assuming that only
such elementary marginal contributions matter in which an agent, after
leaving a coalition, becomes a singleton. Conversely, Bolger assigned
equal importance to all elementary marginal contributions and Macho-Stadler et al. assumed that this importance depends on the size of the coalition the agent joins.

Of course, these combinations represent just a few of the ways in
which they can be combined together. Since our aim is to accommodate all 
such possibilities, we will use in this paper a general definition of the marginal
contribution in games with externalities introduced by Fujinaka~\cite{Fujinaka04}. Intuitively, in this definition, the marginal contribution of $i$ is a \emph{weighted average}
of all possible elementary marginal contributions of $i$.  Let
$\alpha_i$ denote the weight of $i$, and let $\alpha \defeq \langle
\alpha_i \rangle_{i \in N}$.  Then:
$$[mc^{\alpha}_i(v)](S, P) \defeq \sum_{T \in P_{-S}} \alpha_i(S_{-i}, \tau_i^T(P)) \cdot [v(S, P) - v(S_{-i}, \tau_i^T(P))].$$
Now, for weights $\alpha$ to be well defined, it is required that they are
non negative\footnote{\footnotesize Otherwise, agents with only positive (negative) elementary marginal contributions could be assigned negative (positive) payoff.}, and that they do
not depend on the agents' names (to satisfy symmetry). Finally, for normalization, it is assumed that their sum equals one for every embedded coalition. Formally, weights $\alpha_i: \{(S, P) \in EC \mid i \not \in S\} \rightarrow [0,
1]$ satisfy:
\begin{itemize}
\item $\alpha_i(S, P) = \alpha_{\pi(i)}(\pi(S), \pi(P))$ for every permutation $\pi \in \Omega(N)$ and $(S, P) \in EC$ such that $i \not \in S$;
\item $\sum_{T \in P_{-S}} \alpha_i(S_{-i}, \tau_i^T(P)) = 1$ for every $(S, P) \in EC$ such that $i \in S$.
\end{itemize}

Weights $\alpha$ have a natural interpretation: they represent a probability that a transfer of $i$ from $S$ to another coalition in $P$ takes place. Obviously, all the definitions of marginal contributions proposed in the literature fall within this general definition \cite{Bolger89,PhamDoNorde07,MachoStadlerEtAl07,HuYang10,Skibski11}. The definition of marginal contribution corresponding to a given weighting $\alpha$ will be called $\alpha$-marginality and Null-Player Axiom based on $\alpha$-marginality will be denoted Null-Player Axiom$^{\alpha}$.


\section{Uniqueness}\label{section:uniqeness}
In this section, for every weighting $\alpha$, we define a corresponding extension of the Shapley value to games with externalities, denoted $\varphi^\alpha$, and show that it is the unique solution that meets Shapley's four original axioms. The
intuition for $\varphi^\alpha$ is similar to that for the original Shapley value. Specifically, assume that the agents leave a certain meeting point in a
random order and divide themselves into groups (i.e., coalitions)
outside. The agent that leaves joins any
of the groups outside or creates a new group, each with a certain probability. By doing so, he decreases
the value of the coalition in the meeting point (embedded in the
partition of other agents) by his elementary marginal
contribution. Now, the payoff of this agent is the expected value of
his elementary marginal contributions over all permutations and
partitions.

\begin{theorem}\label{theorem:uniqueness} There exists a unique value
  $\varphi^{\alpha}$ that satisfies Efficiency, Symmetry, Additivity
  and Null-Player Axiom$^{\alpha}$ for every $\alpha$. Moreover, 
the value $\varphi^{\alpha}$ satisfies the following formula:
\begin{equation}\label{eq:the_formula}
\varphi^{\alpha}_i(v) = \frac{1}{|N|!}\hspace{-0.1cm} \sum_{\pi \in \Omega(N)} \sum_{P \in \P} pr^{\alpha}_{\pi}(P) \cdot [v(C_i^{\pi} \cup \{i\}, P_{[C_i^{\pi} \cup \{i\}]}) - v(C_i^{\pi}, P_{[C_i^{\pi}]})]
\end{equation}

\noindent where $P_{[S]} \defeq \{T \setminus S \mid T \in P\} \cup \{S\}$ denotes the partition obtained from $P$ by transferring all $i \in S$ to a new coalition and $pr^{\alpha}_{\pi}(P) \defeq \prod_{i \in N} \alpha_{i}(C_i^{\pi}, P_{[C_i^{\pi}]})$.
\end{theorem}
\noindent \textit{Proof:} The above theorem can be proved either directly from the Shapley's original axioms or indirectly by showing the equivalence between Shapley's and Young's axiomatizations for games with externalities and invoking Fujinaka's theorem of uniqueness for Young's axiomatization. We chose the direct proof and construct it as follows. First we prove that $\varphi^{\alpha}$ satisfies all four axioms. Then, we show that this is the only such value.

\noindent \textbf{Part 1:} We will examine axioms one by one.
\begin{itemize}
\item Efficiency: for any permutation $\pi$ and partition $P$, the elementary marginal contributions add up to $v(N, \{N, \emptyset\})$; thus:
\begin{eqnarray*}
\sum_{i \in N} \varphi^{\alpha}_i(v) & = & \frac{1}{|N|!} \sum_{\pi \in \Omega(N)} \sum_{P \in \P} pr^{\alpha}_{\pi}(P) \cdot \sum_{i \in N} [v(C_i^{\pi} \cup \{i\}, P_{[C_i^{\pi} \cup \{i\}]}) - v(C_i^{\pi}, P_{[C_i^{\pi}]})] \\
& = & \frac{1}{|N|!} \sum_{\pi \in \Omega(N)} \sum_{P \in \P} pr^{\alpha}_{\pi}(P) \cdot v(N, \{N, \emptyset\}) = v(N, \{N, \emptyset\}),
\end{eqnarray*}

\noindent where the last transformation comes from the fact that $pr^{\alpha}_{\pi}(P)$ represents the probability that $P$ will form and this probability depends on order of agents $\pi$. Thus, $\sum_{P \in \P} pr^{\alpha}_{\pi}(P) = 1$ for every permutation $\pi$.
\item Symmetry: formula (\ref{eq:the_formula}) does not favor any player, hence permutation of coalitions' values will permute payoffs accordingly.
\item Additivity: formula (\ref{eq:the_formula}) is clearly additive as $\varphi^{\alpha}_i(v_1 + v_2)$ can be split into two expressions representing $\varphi^{\alpha}_i(v_1)$ and $\varphi^{\alpha}_i(v_2)$.
\item Null-Player Axiom$^{\alpha}$: to see that
$\varphi^{\alpha}$ satisfies the Null-Player Axiom$^{\alpha}$ we will calculate the probability of a given elementary marginal contribution $v(S, P) - v(S_{-i}, \tau_i^T(P))$ using formula (\ref{eq:the_formula}). A transfer from $(S, P)$ will occur only in permutations, where agents from $N \setminus S$ (and only them) leave before agent $i$. For one permutation of agents from $N \setminus S$, i.e., $\pi \in \Omega(N \setminus S)$, the probability that $(S, P)$ will form equals $pr^{\alpha}_{\pi}(S, P) = \prod_{j \in N_{-S}} \alpha_{j}(C_j^{\pi}, P_{[C_j^{\pi}]})$. Moreover, the probability of transfer to $T$ equals $\alpha_i(S_{-i}, \tau_i^T(P))$. Finally, the permutation and arrangement of the remaining players does not have an impact on the value (there are $(|S|-1)!$ such permutations). Now, if we collect all transfers from a given embedded coalition $(S, P)$ we get the following formula:
$$\varphi_i^\alpha(v) \defeq \!\!\!\! \sum_{(S, P) \in EC, i \in S} \!\!\!\! \frac{(|S|-1)!}{|N|!} \!\! \sum_{\pi \in \Omega(N_{-S})} \!\! pr^{\alpha}_{\pi}(S, P) \cdot [mc_i^{\alpha}(v)](S, P).$$
\end{itemize}

\noindent \textbf{Part 2:} Now, we will show that $\varphi^{\alpha}$ is the only value which satisfies all four of Shapley's original axioms. To this end, let us introduce the \emph{class of simple games} $\langle e^{(S, P)} \rangle_{(S, P) \in EC}$:
$$e^{(S,P)}(\tilde{S},\tilde{P}) \defeq
\begin{cases}
1 & \quad \text{if} (S, P) = (\tilde{S}, \tilde{P}) \text{,}\\
0 & \quad \text{otherwise.}\\
\end{cases}$$
This class forms a basis of the game space, i.e., every game can be defined as a linear combination of games $e^{(S, P)}$: $v = \sum_{(S, P) \in EC} v(S, P) \cdot e^{(S, P)}$. Based on Additivity, we have $\varphi(v) = \sum_{(S, P) \in EC} \varphi(v(S, P) \cdot e^{(S, P)})$; thus, it is enough to prove that axioms imply a unique value in simple game $e^{(S, P)}$ (multiplied by a scalar). For this purpose, we will use the reverse induction on the size of $S$, i.e., we will show that the value of game $e^{(S, P)}$ can be calculated from values of simple games for bigger coalitions: $e^{(\tilde{S}, \tilde{P})}$ where $|\tilde{S}| > |S|$. Our base case when $|S| = |N|$ comes from the Efficiency and Symmetry: $\varphi_i(c \cdot e^{(N, \{N, \emptyset\})}) = \frac{c}{|N|}$ for every $i$.

First, let $(S, P)$ be any embedded coalition and assume that $i \not \in S$. Let us consider game $\tilde{v}$ combined from two simple games:
$$\tilde{v} = c \cdot [\alpha_i(S, P) \cdot e^{(S_{+i}, \tau_i^S(P))} + e^{(S, P)}]$$

\noindent It is easy to observe that agent $i$'s marginal contribution to $(S_{+i}, \tau_i^S(P))$ equals zero, as with all other marginal contributions. Thus, from Null-Player Axiom $\varphi_i(\tilde{v}) = 0$ and from Additivity:
\begin{equation}\label{equation:proof_iins}
\varphi_i(c \cdot e^{(S, P)}) = - \varphi_i(c \cdot \alpha_i(S, P) \cdot e^{(S_{+i}, \tau_i^S(P))})
\end{equation}

Now, let us assume otherwise, that $i \in S$ and $|S| < |N|$ (we already considered simple game $e^{(N, \{N, \emptyset\})}$). 
From Efficiency, we have that $v(N, \{N, \emptyset\}) = 0$. Thus, we can evaluate the sum of payoffs of agents from $S$ as the opposite number to the sum of payoffs of outside agents ($ - \sum_{j \not \in S} \varphi_j(c \cdot e^{(S, P)})$). This sum, in turn, can be calculated with formula (\ref{equation:proof_iins}). Now, based on the Symmetry, all agents from $S$ divide their joint payoff equally.
\begin{eqnarray*}
\varphi_i(c \cdot e^{(S, P)}) & = & \frac{1}{|S|} \sum_{k \in S} \varphi_k(c \cdot e^{(S, P)}) = - \frac{1}{|S|} \sum_{j \not \in S} \varphi_j(c \cdot e^{(S, P)}) \\
& = & \frac{1}{|S|} \sum_{j \not \in S} \varphi_j(c \cdot \alpha_j(S, P) \cdot e^{(S_{+j}, \tau_j^S(P)})
\end{eqnarray*}
\normalsize

\noindent Thus, we provided two recursive equations for $\varphi_i(c \cdot e^{(S, P)})$ for both cases: $i \in S$ and $i \not \in S$. This concludes our proof.$\square$

Since formula (\ref{eq:the_formula}) is the same as the one derived by Fujinaka~\cite{Fujinaka04} based on the Young's monotonicity axiomatization, we conclude that both axiomatizations are equivalent:

\begin{corollary}\label{corollary:fujinaka}
Value $\varphi$ satisfies Shapley's axiomatization (Efficiency, Symmetry, Additivity and Null-Player Axiom$^{\alpha}$) if and only if it satisfies Young's axiomatization (Efficiency, Symmetry and Marginality Axiom$^{\alpha}$).\footnote{\footnotesize This axiom is yet another way of introducing marginality to the axiomatization.}
\end{corollary}

\section{An Approximation Algorithm}\label{section:algorithm}
We now present our algorithm for approximating the extended Shapley value for any weighting $\alpha$.
We will use the following sampling process. Let the population be the set of pairs $(\pi, P) \in \Omega(N) \times \P$. In one sample, given permutation $\pi$ and partition $P$, we will measure for each agent $i$ his elementary marginal contribution.
As visible in the formula for $\varphi^{\alpha}$ in Theorem \ref{theorem:uniqueness}, elementary marginal contributions do not occur with the same probability. Thus, to obtain an unbiased estimate we will use \emph{probability sampling} with the odds of selecting a given sample $(\pi, P)$ equal $pr^{\alpha}_{\pi}(P)/|N|!$. To this end, we will select a random permutation (each with equal probability: $1/|N|!$) and then select a partition with probability $pr^{\alpha}_{\pi}(P)$.\footnote{\footnotesize Recall that, intuitively, $pr^{\alpha}_{\pi}(P)$ represents the probability that $P$ will form if players leave the meeting point in order~$\pi$.}
It is important to note that this probability depends on the definition of marginality (hence the $\alpha$ in the superscript) and the difficulty of the sampling process may vary depending on the definition adopted. We will address this issue later.

The pseudocode of this procedure is presented in Algorithm \ref{algorithm:ExtendedShapleyValueAppr}. Our procedure, which approximates $\varphi^{\alpha}$, is parametrized by the game $v$ and number of samples $m$. We will discuss the required number of samples at the end of this section. The main for-loop sums samples elementary marginal contributions (variable $\hat{\varphi}^{\alpha}$). At the end, this sum is divided by the number of samples. To compute the players' contribution we reverse the process of creating partition $P$ from the grand coalition (according to the intuition outlined before): we sequentially transfer players to the new (empty at start) coalition that represents a meeting point and measure the change of its value. 

\begin{algorithm}[t]
\caption{Approximation of $\varphi^{\alpha}$}
\label{algorithm:ExtendedShapleyValueAppr}
\begin{algorithmic}[1]
\ForAllLine{$i \in N$}{$\hat{\varphi}^{\alpha}_i \leftarrow 0$;}

\For{$i \leftarrow 1$ \textbf{to} $m$} 
\State $\pi \leftarrow \textit{ random permutation from } \Omega(N)$;
\State $P \leftarrow \textit{ random partition from } \P \textit{ with probability } pr^{\alpha}_{\pi}(P)$;
\State $S \leftarrow \emptyset$;
\For{$j \leftarrow |N|$ \textbf{downto} $1$}
\State $v_{before} \leftarrow v(S, P)$;
\State \textit{transfer player $\pi^{-1}(j)$ in $P$ to $S$};
\State $v_{after} \leftarrow v(S, P)$;
\State $\hat{\varphi}^{\alpha}_{\pi(j)} \leftarrow \hat{\varphi}^{\alpha}_{\pi(j)} + v_{after} - v_{before}$;
\EndFor

\EndFor

\ForAllLine{$i \in N$}{$\hat{\varphi}^{\alpha}_i \leftarrow \hat{\varphi}^{\alpha}_i / m$;} \\
\Return $\hat{\varphi}^{\alpha}$;
\end{algorithmic}
\end{algorithm}

Now, let us focus on the randomized part of our algorithm (lines 3-4). We generate a random permutation using a well-known Knuth shuffle \cite{Durstenfeld:1964}. As mentioned before, the selection of a partition depends on the definition of $\alpha$ weights. That is why we are able to approximate all (marginality-based) extensions of the Shapley value and, in particular, those already proposed in the literature:

\vspace*{1ex}\noindent\textbf{Externality-free value:}
In the simplest concept of \emph{externality-free value}, the whole probability is assigned to creation of a new coalition \cite{PhamDoNorde07}:
$$\alpha^{free}_i(S, P) \defeq 1 \text{ if } \{i\} \in P \text{,}$$
\noindent and $\alpha^{free}_i(S, P) \defeq 0$ otherwise. 
Thus, the only partition with non-zero probability is the partition of singletons: $P = \{\{i\} \mid i \in N\}$ and selection of $P$ is straightforward.

\vspace*{1ex}\noindent\textbf{Full-of-externalities value:}
To obtain McQuillin's \cite{McQuillin09} value, Skibski \cite{Skibski11} used the marginality that complements the previous one:\footnote{The second condition $P = \{N_{-i}, \{i\}\}$ is a special case in which creating a new coalition has non-zero probability. This comes from the fact that creating a new coalition is the agent's only option.}
$$\alpha^{full}_i(S, P) \defeq \frac{1}{|P|-1} \text{ if } \{i\} \not \in P \text{ or } P = \{N_{-i}, \{i\}\} \text{,}$$
and $\alpha^{full}_i(S, P) \defeq 0$ otherwise.
Again, the random selection simplifies to generating one specific partition, as only the grand coalition $P = \{N\}$ has non-zero probability.

\vspace*{1ex}\noindent\textbf{Bolger value:}
Bolger \cite{Bolger89} was first to propose axiomatization based on the marginality principle. In his definition of marginal contribution every transfer is equally likely:
$$\alpha^{B}_i(S, P) \defeq \frac{1}{|P_{-i}|}.$$
The probability of partition $pr^{\alpha^B}_{\pi}(P)$ depends on the order in which the agents leave.\footnote{For example, consider $pr^{\alpha^B}_{\pi}(\{N_{-i}, \{i\}\})$. If $i$ is the last agent in permutation $\pi$ then $pr^{\alpha^B}_{\pi}(\{N_{-i}, \{i\}\}) = \frac{1}{1} \frac{1}{2} \frac{1}{2} \cdots \frac{1}{2} = \frac{1}{2^{|N|-1}}$. On the other hand, if it is the first one: $pr^{\alpha^B}_{\pi}(\{N_{-i}, \{i\}\}) = \frac{1}{1} \frac{1}{2} \frac{1}{3} \cdots \frac{1}{3} = \frac{1}{2 \cdot 3^{|N|-2}}$.} To select a partition with adequate probability, we simulate the process of leaving as follows: we take agents from the permutation one by one and uniformly select one of the existing coalitions to join or a new one to create.

\vspace*{1ex}\noindent\textbf{Macho-Stadler et al. value:}
In the value proposed by Macho-Stadler et al. \cite{MachoStadlerEtAl07}, the weights of the transfer depend on the size of coalitions:
$$\alpha^{MSt}_i(S, P) \defeq \frac{|P(i)|-1}{|N| - |S|} \text{ if } |P(i)| > 1 \text{,}$$
and $\alpha^{MSt}_i(S, P) \defeq \frac{1}{|N| - |S|}$ otherwise, where $P(i)$ denotes coalition with agent $i$ in $P$. Thus, the numerator is equal to the size of the coalition that agent joined (or $1$ when this coalition is empty).\footnote{We note that these weights correspond to the probabilities in Chinese restaurant process \cite{Aldous85}, well known in probability theory.} Our approach to generate a random partition comes from the observation that probability $pr^{\alpha^{MSt}}_{\pi}(P) = \frac{\prod_{T \in P} |T-1|!}{|N|!}$ equals the odds that a given partition will occur from the decomposition into disjoint cycles of a randomly selected permutation. Thus, we can again generate a random permutation with Knuth shuffle and divide players into coalitions according to the cycles of this permutation.

\begin{figure}[t]
\centerline{\includegraphics[natheight=8.8cm, natwidth=8cm, height=7.1cm, width=13.5cm]{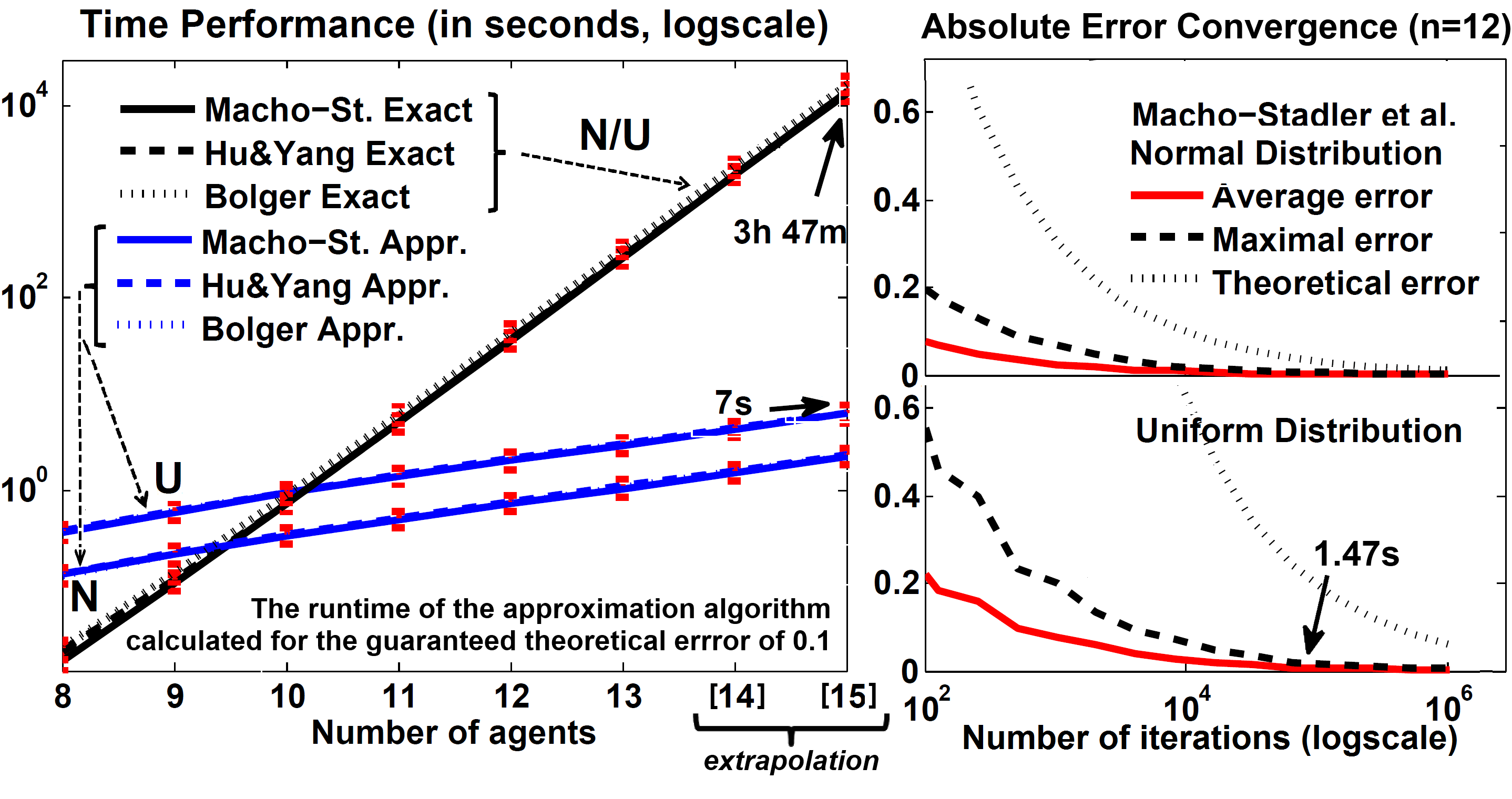}}
\caption{Performance evaluation of Algorithm
\ref{algorithm:ExtendedShapleyValueAppr}.}
\label{fig:Performance_evaluation}
\end{figure}

\vspace*{1ex}\noindent\textbf{Hu and Yang value:}
Hu and Yang \cite{HuYang10} designed their value in such a way that probabilities of every partition are equal: $pr^{\alpha^{HY}}_{\pi}(P) = \frac{1}{|P|}$. Thus,
$$\alpha^{HY}_i(S, P) \defeq \frac{|\{\tilde{P} \in \P \mid \tilde{P}_{[S]} = P\}|}{|\{\tilde{P} \in \P \mid \tilde{P}_{[S \cup \{i\}]} = P\}|}\text{.}$$
To generate a random partition we introduce the following technique.
We will create the partition successively, for each agent selecting the coalition to join with probability that corresponds to the number of partitions of $N$ that cover (i.e., respect) the obtained partial partition. For example, player $2$ will join agent $1$ with probability $\frac{Bell(n-1)}{Bell(n)}$. This is because both agents appear in $Bell(n-1)$ of $Bell(n)$ partitions together. Now, the number of partitions of $N$ that cover given partial partition $P_k$ of $k$ agents depends only on the number of coalitions in $P_k$. Thus, all the probabilities of transfers to the existing coalitions are the same. Based on this analysis, we will first randomly decide whether the player forms a new coalition and if not, we will pick any existing coalition (all with the same probability). We note here that the probability of creating a new coalition by player $k+1$ entering partition $P_k$ can be precalculated, i.e., calculated once, before the sampling. This can be done in $\mathcal{O}(n^2)$ time using dynamic programing.
What is also important from the computational point of view, this ratio is not less that $\frac{1}{|N|}$, thus we avoid precision problems that arises in other methods proposed in the literature \cite{Stam83}.

\begin{table*}[t]
\setlength{\tabcolsep}{0.8pt}
\centering
	\begin{tabular}{lc|cccc|cccccccccccccccccc}
 & & \footnotesize EFF & \footnotesize SYM & \footnotesize ADD & \footnotesize NP$^{\alpha}$ & \footnotesize MARG$^{\alpha}$ & \footnotesize LIN & \footnotesize WNP & \footnotesize \tworows{SS}{SI} & \footnotesize WM & \footnotesize \tworows{GEN}{REC} & \footnotesize CAR & \footnotesize \tworows{OLG}{ECA} & \footnotesize \tworows{FEFF}{ECS} & \footnotesize \tworows{ECAD}{ECNP} & \footnotesize \tworows{CPO}{LEFF} & \footnotesize \tworows{OW}{CNS} & \footnotesize \tworows{SCN}{SCS} & \footnotesize ME\\[1ex]
\hline
\multicolumn{20}{l}{\sc{Marginal values - \textbf{all can be derived from Shapley's original axioms (Theorem~\ref{theorem:uniqueness})}}} \\
\hline
\multicolumn{10}{l}{\sc{\mbox{\ \ \ \ \ \ } \small (a) axiomatizations based on marginality}} \\
\hline
\small Pham Do \& Norde & \footnotesize 2007 & $\times$ & $\times$ & $\times$ & $\circ$\\
\small Hu \& Yang & \footnotesize 2009 & $\times$ & $\times$ & $\times$ & $\circ$\\
\small Skibski & \footnotesize 2011 & $\times$ & $\times$ & $\times$ & $\circ$ & & $\times$\\
\small Bolger & \footnotesize 1989 & $\times$ & $\times$ & $\times$ & & $\circ$ & $\times$ & $\times$\\
\small De Clippel \& Serrano & \footnotesize 2008 & $\times$ & $\times$ & & & $\circ$\\
\hline
\multicolumn{10}{l}{\sc{\mbox{\ \ \ \ \ \ } \small (b) non-marginality axiomatizations}} \\
\hline
\small Macho-Stadler et al. & \footnotesize 2007 & $\times$ & $\times$ & $\times$ & & & $\times$ & $\times$ & $\times$\\
\small McQuillin & \footnotesize 2008 & $\times$ & $\times$ & $\times$ & & & $\times$ & $\times$ & & $\times$ & $\times$\\
\hline
\hline
\multicolumn{10}{l}{\sc{Other values}} \\
\hline
\small Myerson & \footnotesize 1977 & & $\times$ & $\times$ & & & & & & & & $\times$\\
\small Albizuri et al. & \footnotesize 2005 & $\times$ & $\times$ & $\times$ & & & & & & & & & $\times$\\
\small Hafalir & \footnotesize 2007 & & & & & & & & & & & & & $\times$ & $\times$\\
\small Maskin & \footnotesize 2007 & & & & & & & & & & & & & & & $\times$ & $\times$\\
\small Grabisch \& Funaki & \footnotesize 2012 & $\times$ & & $\times$ & & & $\times$ & & & & & & & & & & & $\times$ & $\times$\\
\hline
\end{tabular}
\caption{\small
Existing axiomatizations (that guarantee uniqueness) for various extensions of Shapley value to games with externalities ($\times$ denotes the axiom used and $\circ$ denotes a special case of the axiom). Axioms:
Efficiency (EFF), Symmetry (SYM), Additivity (ADD), Null-Player Axiom$^{\alpha}$ (NP$^{\alpha}$), Marginality$^{\alpha}$ (MARG$^{\alpha}$), Linearity (LIN),
Weak Null-player Axiom (WNP), Strong Symmetry (SS), Similar Influence (SI), Weak Monotonicity (WM), Rules Of Generalization (GEN), Recursion (REC), Carrier (CAR), Oligarchy Axiom (OLG), Embedded Coalition Anonymity (ECA), Fully Efficient (FEFF), Efficient-Cover Anonymous (ECS), Efficient-Cover Additive (ECAD), Efficient-Cover Null-player Axiom (ECNP), Coalition Pareto Optimality (CPO), Limited Efficiency (LEFF), Opportunity Wages (OW), Consistency (CNS), Null-Player Axiom for the Scenario-Value (SCN), Symmetry for the Scenario-Value (SCS), Markovian and Ergodic Axiom (ME)}
\label{table:extendedshapleyvalues}
\end{table*}

\subsection{Error analysis}
Let us briefly discuss the number of samples needed to obtain a required precision of the result. It is clear from the Theorem \ref{theorem:uniqueness} that the estimator is unbiased: $E[\hat{\varphi}_i^{\alpha}] = \varphi_i^{\alpha}$. The variance equals $V[\hat{\varphi}_i^{\alpha}] = \frac{\sigma^2}{m}$ where $m$ is the number of samples and $\sigma^2$ is the variance of the population:
\begin{eqnarray*}
\sigma^2 & = & \frac{1}{|N|!} \sum_{\pi \in \Omega(N)} \sum_{P \in \P} pr^{\alpha}_{\pi}(P) \cdot ([v(C_i^{\pi} \cup \{i\}, P_{[C_i^{\pi} \cup \{i\}]}) - v(C_i^{\pi}, P_{[C_i^{\pi}]})] - \varphi_i^{\alpha}(v))^2.
\end{eqnarray*}
Now, based on the central limit theorem, $\hat{\varphi}_i^{\alpha} \sim N(\varphi_i^{\alpha}, \frac{\sigma^2}{m})$. Assume we want to obtain an error not bigger than $\epsilon$ with the probability not smaller than $1-\beta$, i.e., we need to satisfy the following inequality: $P(|\hat{\varphi}_i^{\alpha} - \varphi_i^{\alpha}| \le \epsilon) \ge 1-\beta$. But $P(|\hat{\varphi}_i^{\alpha} - \varphi_i^{\alpha}| \le \epsilon) = \Phi(\frac{\epsilon \cdot \sqrt{m}}{\sigma}) - \Phi(\frac{- \epsilon \cdot \sqrt{m}}{\sigma}) = 2 \cdot \Phi(\frac{\epsilon \cdot \sqrt{m}}{\sigma}) - 1$ where $\Phi(x)$ is the cumulative distribution function of the standard normal distribution. Therefore, $\Phi(\frac{\epsilon \cdot \sqrt{m}}{\sigma}) \ge 1-\frac{\beta}{2}$ and finally:
$m \ge \frac{\sigma^2}{\epsilon^2} \cdot (\Phi^{-1}(1-\frac{\beta}{2}))^2,$
where $\Phi^{-1}(x)$ is the quantile function, i.e., $P(X \ge \Phi^{-1}(x)) = x$ for $X \sim N(0, 1)$. For example, for the uncertainty $\beta = 0.01$ holds $\Phi^{-1}(0.995) \approx 2.57$.

Next, we need to find an upper bound for $\sigma^2$. To this end, following Castro et al. \cite{Castro09}, we will assume that we know some limits $min_i,max_i$ on the player's marginal contribution, i.e., $min_i \le v(C_i^{\pi} \cup \{i\}, P_{[C_i^{\pi} \cup \{i\}]}) - v(C_i^{\pi}, P_{[C_i^{\pi}]}) \le max_i$ for every $\pi \in \Omega(N)$ and $P \in \P$.
Then, the $\sigma^2$ is maximized when all marginal contributions equal $min_i$ or $max_i$, and the average equals $\frac{min_i+max_i}{2}$ (to achieve this, the sum of probabilities of maximal marginal contributions must equal the sum of probabilities of minimal marginal contributions). Finally, $\sigma^2 \le \frac{(max_i-min_i)^2}{4}$.

\section{Performance evaluation}


We test our algorithm on two distributions popular in the literature on coalitional games \cite{LarsonSandholm00}:
\begin{itemize}
\item normal: $v(S, P) = |S| \cdot N(1, 0.1)$; here, the bounds are: $min_i = - n \cdot 0.6 + 1.3$ and $max_i = n \cdot 0.6 + 0.7$.
\item uniform: $v(S, P) = |S| \cdot U(0, 1)$; here, in a extreme case, $min_i = -n+1$ and $max_i = n$.
\end{itemize}
\noindent where, in the case of the normal distribution, we place the following additional limits: $0.7 \le N(1, 0.1) \le 1.3$.\footnote{These instances only happen with probability $0.14\%$.}

Figure \ref{fig:Performance_evaluation}(a) presents the time performance of our algorithm for $n = 8, 9, \ldots, 15$ and compares it to the exact brute-force approach---the only known alternative.\footnote{The simulations run on a PC-i7, 3.4 GHz and 8 GB of RAM.} For $n=11$ our approximation algorithm outperforms the exact brute-force at the first time. Already for $n = 15$, it would take almost $4$ hours to compute the exact output (extrapolated result), whereas our algorithm returns the approximated solution in less than $7$ seconds (with the guaranteed error of $0.1$). 

Figure \ref{fig:Performance_evaluation}(b) shows that the maximal error obtained from the random game is a few times lower than the theoretical error (for both distributions). For instance, for game of $12$ players, which takes the brute-force algorithm more than $37$ seconds to calculate, the maximal error of $0.018$ is obtained after $1.47$ seconds ($65$K samples). Moreover, the error clearly tends to zero, which shows that our estimator is indeed unbiased.



\section{Conclusions and related work}\label{section:related_work}
We proved in this paper that, for any well-defined marginal contribution, Shapley's original four axioms imply a  unique value for games with externalities. Our theorem covers all previous values that can be based on marginality. Building upon this we develop the first approximation algorithm to compute such values and evaluate its performance.

We summarize the extensions of Shapley value to games with
externalities in Table \ref{table:extendedshapleyvalues}. 
First, four authors proposed their own definitions of marginality, a subcases of $\alpha$-marginality, and proved uniqueness based on the marginality axiomatizations \cite{PhamDoNorde07,HuYang10,Skibski11,Bolger89}. To this end, some strengthen standard axioms: Skibski added Linearity to the standard Shapley's axiomatization; Bolger used Young's axiomatization, but added Linearity and the Weak Null-Player Axiom. De Clippel and Serrano~\cite{ClippelSerrano08} studied externality-free marginality proposed by Pham Do \& Norde applied to Young's axiomatization.
All above uniqueness results are special cases of Theorem~\ref{theorem:uniqueness}. 
Furthermore, Macho-Stadler et al. \cite{MachoStadlerEtAl07} and McQuillin \cite{McQuillin09} proposed new values based on more complex axiomatizations. Although their axiomatizations
move away from the concept of marginality both of these values can
be obtained using marginality axiomatization.  Other authors used
axiomatizations which conflict with Shapley's understanding of
fairness: Myerson's \cite{Myerson77} value may assign negative payoff to a player from
the only non-zero coalition in a game; Albizuri et al.'s \cite{albizuri2005axiom} and Grabish-Funaki's \cite{GrabischFunaki08}
values grant players with no strength non-zero payoffs; and Hafalir \cite{Hafalir07} and Maskin \cite{Maskin03} questioned that the grand coalition forms and declined Efficiency. 

Finally, we mention the relevant computer science literature that includes works on concise representations of games with externalities (\textit{e.g.} \cite{Ichimura:et:al:2011,Michalak10ecai}) and approximation algorithms for games with no externalities (\cite{Fatima_2007,Bachrach_2008}).


\bibliographystyle{abbrv}
\bibliography{approximation_technical_report}

\begin{thebibliography}{10}

\bibitem{albizuri2005axiom}
M.~Albizuri, J.~Arin, and J.~Rubio.
\newblock An axiom system for a value for games in partition function form.
\newblock {\em Int. Game Theory Review}, 7(01):63--72, 2005.

\bibitem{Aldous85}
D.~Aldous.
\newblock Exchangeability and related topics.
\newblock In {\em Ecole d'Ete de Probabilities de Saint-Flour XIII 1983}, pages
  1--198. 1985.

\bibitem{Bachrach_2008}
Y.~Bachrach, E.~Markakis, A.~Procaccia, J.~Rosenschein, and A.~Saberi.
\newblock Approximating power indices.
\newblock In {\em AAMAS '08: Proceedings of the 7th International Joint
  Conference on Autonomous Agents and Multi-Agent Systems}, pages 943--950,
  2008.

\bibitem{Banerjee:Landon}
B.~Banerjee and L.~Kraemer.
\newblock Coalition structure generation in multi-agent systems with mixed
  externalities.
\newblock AAMAS '10, pages 175--182, 2010.

\bibitem{Bolger89}
E.~Bolger.
\newblock A set of axioms for a value for partition function games.
\newblock {\em International Journal of Game Theory}, 18(1):37--44, 1989.

\bibitem{Castro09}
J.~Castro, D.~G{\'o}mez, and J.~Tejada.
\newblock Polynomial calculation of the shapley value based on sampling.
\newblock {\em Computers \& OR}, 36(5):1726--1730, 2009.

\bibitem{Chalkiadakis:et:al:2011}
G.~Chalkiadakis, E.~Elkind, and M.~Wooldridge.
\newblock Computational aspects of cooperative game theory.
\newblock {\em Synthesis Lectures on Artificial Intelligence and Machine
  Learning}, 5(6):1--168, 2011.

\bibitem{ClippelSerrano08}
G.~de~Clippel and R.~Serrano.
\newblock Marginal contributions and externalities in the value.
\newblock {\em Econometrica}, 76(6):1413--1436, 2008.

\bibitem{deKeijzer:Schafer:2012}
B.~de~Keijzer and G.~Sch{\"a}fer.
\newblock Finding social optima in congestion games with positive
  externalities.
\newblock In {\em Algorithms -- ESA 2012}, volume 7501 of {\em LNCS}, pages
  395--406. Springer, 2012.

\bibitem{Durstenfeld:1964}
R.~Durstenfeld.
\newblock Algorithm 235: Random permutation.
\newblock {\em Commun. ACM}, 7(7):420, 1964.

\bibitem{Maskin03}
E.~E.~Maskin.
\newblock Bargaining, coalitions, and externalities.
\newblock presidential address of the econometric society, 2003.

\bibitem{Fatima_2007}
S.~Fatima, M.~Wooldridge, and N.~R. Jennings.
\newblock A randomized method for the shapley value for the voting game.
\newblock In {\em AAMAS}, pages 955--962, 2007.

\bibitem{Fujinaka04}
Y.~Fujinaka.
\newblock On the marginality principle in partition function form games.
\newblock Unpublished manuscript, 2004.

\bibitem{GrabischFunaki08}
M.~Grabisch and Y.~Funaki.
\newblock A coalition formation value for games with externalities.
\newblock Technical Report b08076, Université Panthéon-Sorbonne (Paris 1),
  Centre d'Economie de la Sorbonne, 2008.

\bibitem{Hafalir07}
I.~Hafalir.
\newblock Efficiency in coalition games with externalities.
\newblock {\em Games and Economic Behavior}, 61(2):242--258, 2007.

\bibitem{HartMasColell89}
S.~Hart and A.~Mas-Colell.
\newblock Potential, value and consistency.
\newblock {\em Econometrica}, 57:589--614, 1989.

\bibitem{HuYang10}
C.-C. Hu and Y.-Y. Yang.
\newblock An axiomatic characterization of a value for games in partition
  function form.
\newblock {\em SERIEs}, 1(4):475--487, 2010.

\bibitem{Ichimura:et:al:2011}
R.~Ichimura, T.~Hasegawa, S.~Ueda, A.~Iwasaki, and M.~Yokoo.
\newblock Extension of mc-net-based coalition structure generation: handling
  negative rules and externalities.
\newblock In {\em AAMAS}, pages 1173--1174, 2011.

\bibitem{LarsonSandholm00}
K.~Larson and T.~Sandholm.
\newblock Anytime coalition structure generation: An average case study.
\newblock {\em Journal of Experimental and Theoretical AI}, 12:40--47, 2000.

\bibitem{MachoStadlerEtAl07}
I.~Macho-Stadler, D.~Perez-Castrillo, and D.~Wettstein.
\newblock Sharing the surplus: An extension of the shapley value for
  environments with externalities.
\newblock {\em Journal of Economic Theory}, 135(1):339--356, 2007.

\bibitem{McQuillin09}
B.~McQuillin.
\newblock The extended and generalized shapley value: Simultaneous
  consideration of coalitional externalities and coalitional structure.
\newblock {\em Journal of Economic Theory}, 144(2):696--721, 2009.

\bibitem{Michalak10amas}
T.~Michalak, D.~Marciniak, M.~Szamotulski, T.~Rahwan, M.~Wooldridge,
  P.~McBurney, and N.~Jennings.
\newblock A logic-based representation for coalitional games with
  externalities.
\newblock In {\em AAMAS}, pages 125--132, 2010.

\bibitem{Michalak10ecai}
T.~Michalak, T.~Rahwan, D.~Marciniak, M.~Szamotulski, and N.~Jennings.
\newblock Computational aspects of extending the shapley value to coalitional
  games with externalities.
\newblock In {\em ECAI}, pages 197--202, 2010.

\bibitem{Myerson77}
R.~Myerson.
\newblock Values of games in partition function form.
\newblock {\em International Journal of Game Theory}, 6:23--31, 1977.

\bibitem{Myerson80}
R.~Myerson.
\newblock Conference structures and fair allocation rules.
\newblock {\em International Journal of Game Theory}, 9:169--82, 1980.

\bibitem{PhamDoNorde07}
K.~PhamDo and H.~Norde.
\newblock The shapley value for partition function form games.
\newblock {\em International Game Theory Review}, 9(02):353--360, 2007.

\bibitem{Rahwan:et:al:2012}
T.~Rahwan, T.~Michalak, M.~Wooldridge, and N.~Jennings.
\newblock Anytime coaliton structure generation in multi-agent systems with
  positive or negative externalities.
\newblock {\em Artificial Intelligence}, 186:95--122, 2012.

\bibitem{Shapley53}
L.~Shapley.
\newblock A value for n-person games.
\newblock In H.~Kuhn and A.~Tucker, editors, {\em Contributions to the Theory
  of Games, volume II}, pages 307--317. Princeton University Press, 1953.

\bibitem{Skibski11}
O.~Skibski.
\newblock Steady marginality: A uniform approach to shapley value for games
  with externalities.
\newblock In {\em Symposium on Algorithmic Game Theory}, volume 6982 of {\em
  LNCS}, pages 130--142. Springer, 2011.

\bibitem{Stam83}
A.~Stam.
\newblock Generation of a random partition of a finite set by an urn model.
\newblock {\em Journal of Combinatorial Theory, Series A}, 35(2):231--240,
  1983.

\bibitem{ThrallLucas63}
R.~Thrall and W.~Lucas.
\newblock N-person games in partition function form.
\newblock {\em Nav. Res. Logist. Quart.}, 10:281–298, 1963.

\bibitem{Young85}
P.~Young.
\newblock Monotonic solutions of cooperative games.
\newblock {\em International Journal of Game Theory}, 14(2):65--72, 1985.

\end{thebibliography}

\end{document}